\documentclass[onecolumn,amsmath,amssymb,prd,reprint,longbibliography,floatfix,8pt]{revtex4-1}

\usepackage{cancel}
\usepackage{enumitem}
\usepackage[utf8x]{inputenc}
\usepackage{graphicx}
\usepackage{dcolumn}
\usepackage{bm}
\usepackage[switch]{lineno}
\usepackage{float} 
\usepackage{subfigure}
\usepackage{tikz}
\usepackage{multirow}
\usetikzlibrary{arrows.meta}
\usetikzlibrary{arrows,decorations.pathmorphing,backgrounds,positioning,fit,petri}

\begin{document}

\title{Fourth-order quantum master equations reveal that spin-phonon decoherence undercuts long magnetization relaxation times in single-molecule magnets}

\author{Alessandro Lunghi}
\email{lunghia@tcd.ie}
\affiliation{School of Physics, AMBER and CRANN Institute, Trinity College, Dublin 2, Ireland}

\begin{abstract}
{\bf Spin-phonon interaction is known to drive magnetic relaxation in solid-state systems, but little evidence is available on how it affects coherence time. Here we extend fourth-order quantum master equations to account for coherence terms and describe the full effect of up to two-phonon processes on spin dynamics. We numerically implement this method fully ab initio for a single-molecule magnet with large magnetization blocking temperature and show that while strong axial magnetic anisotropy ensures slow magnetic relaxation approaching seconds at 77 K, the superposition of Kramers doublets is coherent for less than 10 ns due to a novel two-phonon pure dephasing mechanism.}
\end{abstract}

\maketitle

\textit{Introduction.} 
Magnetic relaxation studies have been conducted across vastly different length scales, ranging from bulk magnets to nanoparticles, molecules, and single atoms adsorbed on surfaces\cite{gambardella2003giant,donati2016magnetic,sessoli1993magnetic}, showing that the effect of spin-phonon interaction is strongly inhibited when the magnetization experiences a reversal energy barrier introduced by magnetic uniaxial anisotropy. The study of single-molecule magnets (SMMs) has particularly enjoyed large attention in recent years thanks to the possibility of chemically tailoring said magnetic anisotropy, and record-breaking long magnetization relaxation times, $\tau$, have been achieved for coordination compounds based on Dy$^{3+}$ ions with large effective total magnetic moment $J=15/2$ and magnetization reversal barriers exceeding 2000 K\cite{goodwin2017molecular,guo2018magnetic,emerson2025soft}. The application of SMMs in quantum information science has also been pursued due to several advantageous properties of this physical platform with respect to others, including chemical and physical tunability, synthetic reproducibility, and more\cite{atzori2019second,gaita2019molecular}. Early prototypes of SMMs based on clusters of 3d ions had indeed shown that Rabi oscillations and quantum gates can be driven at cryogenic temperatures where relaxation time, $T_1$, is long enough, leaving spin-spin dipolar interactions to limit coherence time, $T_2$, in such conditions\cite{hill2003quantum,ardavan2007will,schlegel2008direct}. However, dipolar contributions to $T_2$ can be mitigated by deuteration or using proton-free environments\cite{zadrozny2015millisecond}, effectively making spin relaxation the hard limit for $T_2$ at all temperatures. More recently, spin coherence in 4f ions\cite{pedersen2016toward,shiddiq2016enhancing}, isotropic spin-1/2 3d ions\cite{bader2014room,zadrozny2015millisecond,atzori2016room} and organic radicals\cite{schafter2023molecular,gorgon2023reversible} has also been explored, with qualitatively similar phenomenology. SMMs with ultra-slow magnetic relaxation are an intriguing candidate for quantum applications, but the limit of $T_2$ posed by spin-phonon coupling in these systems remains unknown, as no experimental or theoretical evidence is available. 

\textit{Quantum master equations.} Here we address this knowledge gap by tapping into the increasingly successful field of ab initio spin dynamics. This method has already provided unique insights into the spin-phonon relaxation of magnetic molecules, ranging from isotropic spin-1/2 systems\cite{mariano2025role} up to strongly correlated 3d metal\cite{lunghi2022toward,mondal2025spin} or lanthanide complexes\cite{goodwin2017molecular,briganti2021complete}, and solid-state paramagnetic defects\cite{mondal2023spin,cambria2023temperature}. A theory of spin-phonon relaxation up to two-phonon contributions has so far been developed\cite{lunghi2022toward,mariano2025role}. However, while second-order contributions have been fully unravelled\cite{lunghi2019phonons,lunghi2020limit}, fourth-order contributions, key to describe low-temperature spin dynamics in SMMs\cite{briganti2021complete}, have so far been determined only for the population terms of the density matrix\cite{lunghi2022toward}. In addition to precluding gaining insights on coherence times at low temperature, this limitation has been shown to potentially lead to catastrophic results in the case of spectral degeneracies able to couple the time evolution of population and coherence terms of the density matrix\cite{lunghi2022toward}. 

In order to access coherence time over the full temperature range, we here set out to derive a 4th-order propagator in Lindblad form able to describe the time evolution of the entire spin density matrix. 
Let us start by describing the system Hamiltonian
\begin{equation}
    \hat{H}= \hat{H}_s + \hat{H}_{ph} + \hat{H}_{sph} \:,
    \label{ham0}
\end{equation}
where $\hat{H}_s$ is the spin Hamiltonian, and $\hat{H}_{ph}$ is the phonon Hamiltonian, which we assume to be harmonic,
\begin{equation}
    \hat{H}_{ph}= \sum_{\alpha} \hbar\omega_{\alpha} ( \hat{n}_\alpha + \frac{1}{2} )\:,
    \label{hamph}
\end{equation}
where $\hbar\omega_{\alpha}$ is the $\alpha$-phonon quantum energy and $\hat{n}_\alpha=a^{\dagger}_{\alpha}a_{\alpha}$ is the number operator written in terms of creation and annihilation operators. Finally, the third term in Eq. \ref{ham0} is the spin-phonon interactions, which in the weak-coupling linear regime reads
\begin{equation}
    \hat{H}_{sph}= \sum_{\alpha} \left( \frac{\partial \hat{H}_s}{\partial q_{\alpha}} \right) q_{\alpha} \:,
\end{equation}
where $q_{\alpha}=2^{-1/2}\left( a^{\dagger}_{\alpha} + a_{\alpha} \right)$ are the phonons' atomic displacements.\\

The time evolution of the spin system under the influence of a weakly coupled phonon bath is here tackled with time-local quantum master equations\cite{timm2011time}, 
\begin{equation}
 \frac{d\hat{\rho}(t)}{dt} =  \hat{\hat{R}}(t,t_0) \: \hat{\rho}(t) \xrightarrow{t_0 \rightarrow -\infty} \hat{\hat{R}} \: \hat{\rho}(t)
\label{tcl}
\end{equation}
where $\hat{\hat{R}}$ is a super-operator describing the effect of the phonon bath over the dynamics of the reduced spin density operator $\hat{\rho}(t)$ expressed in the interaction picture. As depicted in Eq. \ref{tcl}, the superoperator $\hat{\hat{R}}$ becomes time independent in the Markovian limit, i.e. long times after the system has been initialized at $t_0$. The latter approximation is well justified for molecular crystals as phonons relax to equilibrium on much faster timescales than spin\cite{albino2021temperature,nabi2023accurate}. In order to make Eq. \ref{tcl} tractable, the super-operator $\hat{\hat{R}}$ is often expanded perturbatively, $\hat{\hat{R}}= \sum_{n=1}^{\infty}\hat{\hat{R}}^{(2n)}$. Recently, the regularized T-matrix approach\cite{timm2011time} has been used to determine an explicit expression of $\hat{\hat{R}}^{(2)}$ and $\hat{\hat{R}}^{(4)}$ under the Markov and secular approximations\cite{lunghi2019phonons,lunghi2022toward}. By expressing all operators in the eigenspace of $\hat{H}_s$, the population terms of $\hat{\hat{R}}^{(2)}$ read
\begin{equation}
    R^{(2)}_{bb,aa}=\frac{2\pi}{\hbar^2} \sum_{\alpha} | V_{ba}^\alpha  |^2G^{(2)}(\omega, \omega_{\alpha}) \:, 
    \label{R2}
\end{equation}
where $V_{ba}^\alpha=\langle b| \partial \hat{H}_{s} / \partial Q_{\alpha} |a \rangle$, and $\hbar\omega=\hbar\omega_{ba}=E_{b}-E_{a}$. $G^{(2)}$ reads
\begin{equation}
G^{(2)}(\omega, \omega_{\alpha}) = \delta(\omega-\omega_\alpha)\bar{n}_\alpha +\delta(\omega +\omega_\alpha)(\bar{n}_\alpha +1)    \:,
\label{G2}
\end{equation}
where $\bar{n}_\alpha=(e^{\hbar\omega_{\alpha}/\mathrm{k_B}T}-1)^{-1}$ is the Bose-Einstein distribution accounting for the thermal population of phonons, $\mathrm{k_B}$ is the Boltzmann constant, and the Dirac delta functions enforce energy conservation during the absorption and emission of phonons by the spin system, respectively. 
There are three possible two-phonon contributions to $\hat{\hat{R}}^{(4)}$, double phonon emission, double phonon absorption and simultaneous emission-absorption. For the absorption of phonon $\alpha$ and emission of phonon $\beta$, $R^{(4)}_{bb,aa}$ is
\begin{equation}
    R^{(4)}_{bb,aa}  = \frac{2\pi}{\hbar^2} \sum_{\alpha>\beta}\left | T^{\alpha\beta,+}_{ba} + T^{\beta\alpha,-}_{ba} \right|^2G^{(4)} (\omega, \omega_{\alpha}, \omega_{\beta})\:,
    \label{R4}
\end{equation}
where the terms
\begin{equation}
T^{\alpha\beta,\pm}_{ba} = \sum_{c} \frac{ V_{bc}^\alpha V_{ca}^\beta }{E_c -E_a \pm \hbar\omega_\beta} 
\label{R4T}
\end{equation}
describe the contribution of virtual transitions to excited spin states $|c\rangle $, and $G^{(4)}$ reads
\begin{equation}
G^{(4)}(\omega, \omega_{\alpha},\omega_{\beta}) = \delta(\omega-\omega_\alpha+\omega_\beta)\hat{n}_\alpha(\hat{n}_\beta +1).
\label{G4}
\end{equation} 
In order to extend these results to the description of the full reduced spin density matrix, we recall the fundamental theorem by Lindblad\cite{lindblad1976generators} that determines that any process described by a Markovian semi-group leads to a generalized and universal form of quantum master equations, which expressed in the eigenbasis of $\hat{H}_s$ read
\begin{widetext}
\begin{equation}
   \frac{d\hat{\rho}_{ab,cd}(t)}{dt} =  \sum_{\kappa} \gamma_{\kappa} 
   \left[ L_{db}^{\kappa} L_{ac}^{\kappa} - \frac{1}{2} \sum_j \delta_{bd} L_{jc}^{ \kappa} L_{aj}^{\kappa}- \frac{1}{2} \sum_j \delta_{ac} L_{jb}^{ \kappa} L_{dj}^{\kappa} \right ] \hat{\rho}_{cd}(t) \:,
 \label{lind_el}
\end{equation}
\end{widetext}
%
%
where the jump operators $\hat{L}_{\kappa}$ represent the effect of the bath on the reduced density matrix operator, and $\gamma_{\kappa}$ are positive weights determining the relative efficiency of different $\kappa$ bath contributions. A one-to-one map between Eq. \ref{lind_el} and $R^{(2n)}_{ab,cd}$, order by order, emerges. For instance, under the secular approximation, \textit{i.e.} $R_{ab,cd}=0$ unless $\omega_{bd}=\omega_{ac}=\omega$, the second-order jump operators read
\begin{equation}
 \sqrt{\gamma_{\kappa}} L_{db}^{\kappa} = \frac{ \sqrt{ 2\pi \: G^{(2)}(\omega,\omega_\alpha)}}{\hbar} V^\alpha_{db} \:,
 \label{L2}
\end{equation}
where $\kappa=\alpha$.
As one of the central results of this work, here we propose to use a similar inference to derive an explicit mathematical form for the jump operators from $R^{(4)}_{bb,aa}$, leading to
\begin{equation}
\sqrt{\gamma_{\kappa}} L_{db}^{\kappa} = \frac{ \sqrt{ 2\pi \: G^{(4)}(\omega,\omega_\alpha,\omega_\beta)}}{\hbar}
\left [T^{\alpha\beta,+}_{db} + T^{\beta\alpha,-}_{db}  \right] \:,
    \label{L4}
\end{equation}
where $\kappa$ needs now to be interpreted as running over any distinct pair of $(\alpha,\beta)$ phonons.
Eqs. \ref{L2} and \ref{L4} can then be used to populate Eq. \ref{lind_el} and describe the dynamics of the full spin density matrix. 

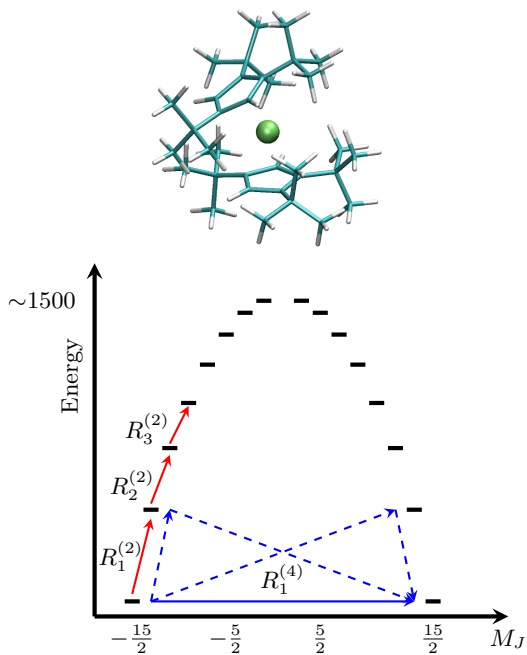
\begin{figure}[h!]
 \begin{center}
    \includegraphics[scale=0.7]{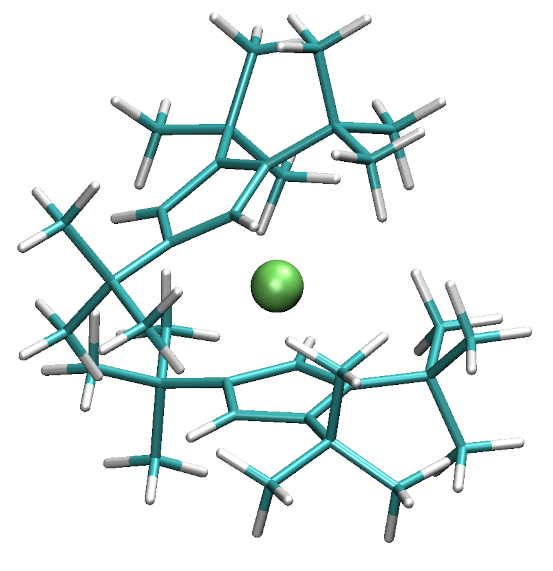} 
    \begin{tikzpicture}     
     
     \node at (1.8,5) {$\sim$1500};
     \node at (8,0.5) {$M_J$};
     
     \draw [ultra thick, ->,>=stealth] (2.5,0.8) -- (2.5,5.5);
     \draw [ultra thick, ->,>=stealth] (2.5,0.8) -- (8,0.8);
     \node [rotate=90] at (2.2,4) {Energy};
         
     \draw [thick,red,->,>=stealth] (3,1.1) -- (3.25,2.1) ;  
     \node at (2.85,1.6) {$R^{(2)}_{1}$};
     
     \draw [thick,red,->,>=stealth] (3.25,2.3) -- (3.5,2.95) ;  
     \node at (3,2.55) {$R^{(2)}_{2}$};;
     
     \draw [thick,red,->,>=stealth] (3.5,3.1) -- (3.75,3.6) ;  
     \node at (3.2,3.3) {$R^{(2)}_{3}$};;

     \draw [thick,blue,->,>=stealth] (3.25,1) -- (6.75,1) ;  
     \draw [thick,dashed,blue,->,>=stealth] (3.25,1) -- (6.5,2.22) ;  
     \draw [thick,dashed,blue,->,>=stealth] (6.5,2.22) -- (6.75,1);  
     \draw [thick,dashed,blue,->,>=stealth] (3.25,1) -- (3.5,2.22) ;  
     \draw [thick,dashed,blue,->,>=stealth] (3.5,2.22) -- (6.75,1);  
     \node at (5,1.3) {$R^{(4)}_{1}$};
     
     \draw [ultra thick] (2.9,1) -- (3.1,1);
     \draw [ultra thick] (3.15,2.22) -- (3.35,2.22);
     \draw [ultra thick] (3.4,3.04) -- (3.6,3.04);
     \draw [ultra thick] (3.65,3.64) -- (3.85,3.64);
     \draw [ultra thick] (3.9,4.15) -- (4.1,4.15);
     \draw [ultra thick] (4.15,4.55) -- (4.35,4.55);
     \draw [ultra thick] (4.4,4.84) -- (4.6,4.84);
     \draw [ultra thick] (4.65,5) -- (4.85,5);
     \draw [ultra thick] (5.15,5) -- (5.35,5);
     \draw [ultra thick] (5.4,4.84) -- (5.6,4.84);
     \draw [ultra thick] (5.65,4.55) -- (5.85,4.55);
     \draw [ultra thick] (5.9,4.15) -- (6.1,4.15);
     \draw [ultra thick] (6.15,3.64) -- (6.35,3.64);
     \draw [ultra thick] (6.4,3.04) -- (6.6,3.04);
     \draw [ultra thick] (6.65,2.22) -- (6.85,2.22);
     \draw [ultra thick] (6.9,1) -- (7.1,1);
     
     \node at (3,0.5) {$-\frac{15}{2}$};
     \node at (4.25,0.5) {$-\frac{5}{2}$};
     \node at (5.5,0.5) {$\frac{5}{2}$};
     \node at (7,0.5) {$\frac{15}{2}$};
    \end{tikzpicture}
\end{center}
 \caption{Top panel: DyCp molecular structure. Dy is represented in acid green, C in green and H in white. Bottom panel: The Orbach relaxation process is depicted with red continuous arrows, each mediated by a second order transition with rate $R^{(2)}_i$. The continuous blue arrow represents the fourth-order transition with rate $R^{(4)}_1$ causing Raman relaxation. The dashed blue arrows show the virtual transitions.}
 \label{Fig1}
\end{figure}

\begin{figure}[h!]
 \begin{center}
    \includegraphics[scale=1]{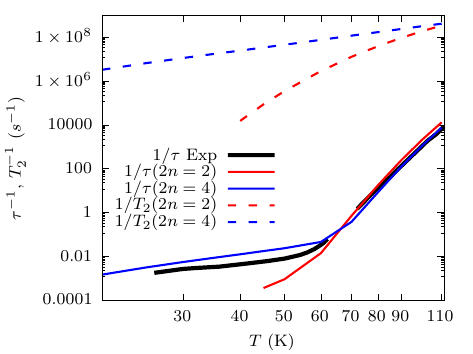}     
\end{center}
 \caption{The experimental magnetization relaxation time\cite{goodwin2017molecular} is reported with a black continuous line. Simulated magnetic relaxation is reported in continuous lines. Simulated decoherence time for the ground state KD is reported with dashed lines. 2nd order results are in red and 4th order ones in blue.}
 \label{Fig2}
\end{figure}

\textit{Numerical Simulations.} We numerically implement the calculation of $R^{(2)}_{ab,cd}$ and $R^{(4)}_{ab,cd}$ for the single-molecule magnet
[DyCp$^{\textrm{ttt}}_2$]$^{+}$ (Cp$^{\textrm{ttt}}$=[C$_{5}$H$_{5} ^{\;\;\textrm{t}}$Bu$_{3}$-1,2,4]), DyCp in short from now on, a coordination complex exhibiting a $J=15/2$ ground state and a record-breaking zero-field splitting of the the $M_J$ components of about 1500 cm$^{-1}$\cite{goodwin2017molecular}, as depicted in Fig. \ref{Fig1}. 
The spin Hamiltonian of this compound can be modelled through a zero-filed splitting term expressed as a combination of tesseral operators and a Zeeman term,
\begin{equation}
    \hat{H}_s= \sum_{l=2}^{6} \sum_{m=-l}^{l} B^l_m \hat{O}^l_m (\vec{\mathbf{J}}) + \mu_\mathrm{B} g_J \vec{\mathbf{J}} \cdot \vec{\mathbf{B}} \:,
\end{equation}
with $l$ restricted to even values to preserve Kramers degeneracy in the absence of magnetic fields, $g_J$ is the Lande factor for $J=15/2$ and $\mu_\mathrm{B}$ is the Bohr's magneton. All simulations are conducted by orienting the ground-state KD's easy axis along $z$. DyCp's crystal phonons, the coefficients $B_{lm}$ and their derivatives, fully define Eq. \ref{ham0} and have been previously computed through a combination of density functional theory and multireference quantum chemistry\cite{lunghi2022toward}. Values of magnetic relaxation rate, $\tau^{-1}$, are extracted from simulations by diagonalizing the generalized matrices $\hat{\hat{R}}^{(2n)}$ and identifying the eigenvalue associated to an eigenvector corresponding to a transition between the states forming the fundamental Kramers Doublet (KD), i.e. the states with maximum and minimum magnetization value. Simulations results, also reported in Fig. \ref{Fig2}, accurately reproduce experiments across the entire available dataset. Above about 60 K, $\hat{\hat{R}}^{(2)}$ determines the relaxation of the magnetization through the Orbach mechanism, which involves subsequent single-phonon transitions to higher energy KDs (schematically depicted in Fig. \ref{Fig1}), and eventually reaching the reversal of the magnetization by emitting single-phonons back to the ground-state KD. At low temperature, the effect of $\hat{\hat{R}}^{(4)}$ leads to two-phonon Raman relaxation, which involves an intra ground-state KD process mediated by virtual transitions to excited KDs and the simultaneous emission and absorption of a pair of phonons whose energy difference matches the Zeeman splitting of the ground-state KD. The simulation of Raman relaxation can now be obtained without the need to break Kramers degeneracy and results are fully rotationally invariant, overcoming previous theoretical limitations\cite{lunghi2022toward}.

\begin{figure}[h!]
 \begin{center}
    \includegraphics[scale=1]{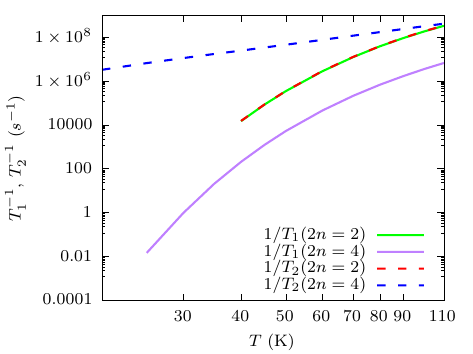}     
\end{center}
 \caption{Second-order ($2n=2$) $T_2$ and $T_1$ are reported in continuous green and dashed red lines, respectively. Fourth-order ($2n=4$) $T_2$ and $T_1$ are reported in continuous purple and dashed blue lines, respectively. The time constants refer to the coherence superposition of the ground-state KD states.}
 \label{Fig3}
\end{figure}

We then turn to the simulation of coherence times, now enabled by the full expression of $\hat{\hat{R}}^{(2)}$ and $\hat{\hat{R}}^{(4)}$. To do this, we chose to study the coherent superposition among the ground-state KD, namely between the states $| a \rangle = | M_J \simeq 15/2 \rangle$ and $| b \rangle = | M_J \simeq -15/2 \rangle$. Fig. \ref{Fig2} reports the values of $T_2$, computed as $T_2^{-1} = -R^{(2n)}_{ab,ab}$. Simulated coherence times are dramatically shorter than the relaxation time of the molecular magnetization. We interpret this results by writing down the expression of $T_2$ for a pair of states. It follows from Eq. \ref{lind_el} that the decoherence rate is due to spin relaxation and pure spin dephasing contributions, $(T_2)^{-1}=(2T_1)^{-1}+(T^*_2)^{-1}$, with 
 \begin{equation}
  \frac{1}{2T_1}=  \sum_{\kappa} \gamma_{\kappa} \left[ \frac{1}{2} \sum_{j\ne a} L^{k}_{ja}L^{k}_{aj} +\frac{1}{2} \sum_{j\ne b} L^{k}_{jb}L^{k}_{bj} \right]   \:, 
 \label{t1}
 \end{equation}
 \begin{equation}
   \frac{1}{T^*_2}= \sum_{\kappa} \gamma_{\kappa} \left[-L_{bb}^{\kappa} L_{aa}^{\kappa} + \frac{1}{2} L_{aa}^{ \kappa} L_{aa}^{\kappa} + \frac{1}{2} L_{bb}^{ \kappa} L_{bb}^{\kappa} \right] \:.
   \label{t2}
 \end{equation}      
The first thing to note is that in multi-level systems, such as DyCp, the magnetic relaxation time and the $T_1$ contribution to $T_2$ for a pair of states might be intrinsically different. For instance, in the high-temperature regime dominated by Orbach relaxation, a series of single-phonon absorptions is necessary to reverse the magnetic moment and cause its relaxation, with the overall rate of the total process diminishing for each additional step taken to climb the magnetization reversal barrier. On the other hand, the $T_1$ for a pair of states is limited by transitions to any state, not necessarily requiring to climb the full magnetization reversal energy barrier, and thus occurring at a much faster rate. This is confirmed by Fig. \ref{Fig3}, which shows that  $T_1$ exponentially decreases with temperature, $A \: \mathrm{exp}(-U/\mathrm{k_B}T)$, with $U$=438.0 cm$^{-1}$ and $U$=466.9 cm$^{-1}$, for second and fourth order, respectively, and thus mediated by phonon-induced transitions to the first excited KD computed at about 451 cm$^{-1}$.

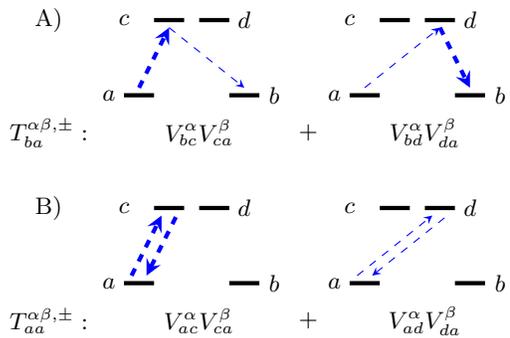
\begin{figure}[h!]
 \begin{center}    
    \begin{tikzpicture}     

     \draw [ultra thick] (1,2.5) -- (1.4,2.5);          
     \draw [ultra thick] (2.4,2.5) -- (2.8,2.5);    
     \draw [ultra thick] (1.4,3.5) -- (1.8,3.5);     
     \draw [ultra thick] (2,3.5) -- (2.4,3.5);     
     \draw [ultra thick,dashed,blue,->,>=stealth] (1.2,2.6) -- (1.6,3.4);  
     \draw [dashed,blue,->,>=stealth] (1.6,3.4) -- (2.6,2.6);  

     \draw [ultra thick] (4,2.5) -- (4.4,2.5);          
     \draw [ultra thick] (5.4,2.5) -- (5.8,2.5);    
     \draw [ultra thick] (4.4,3.5) -- (4.8,3.5);     
     \draw [ultra thick] (5,3.5) -- (5.4,3.5);     
     
     \draw [dashed,blue,->,>=stealth] (4.2,2.6) -- (5.2,3.4);  
     \draw [ultra thick,dashed,blue,->,>=stealth] (5.2,3.4) -- (5.6,2.6);  

     \node at (0.8,2.5) {$a$};
     \node at (3,2.5) {$b$};
     \node at (1,3.5) {$c$};
     \node at (2.6,3.5) {$d$};
     
     \node at (0.8+3,2.5) {$a$};
     \node at (3+3,2.5) {$b$};
     \node at (1+3,3.5) {$c$};
     \node at (2.6+3,3.5) {$d$};

     \node at (0,2) {$T^{\alpha\beta,\pm}_{ba}:$};    
     \node at (2,2) {$V^\alpha_{bc}V^\beta_{ca}$};    
     \node at (3.45,2) {$+$};     
     \node at (5,2) {$V^\alpha_{bd}V^\beta_{da}$};
     
     \draw [ultra thick] (1,2.5-2.5) -- (1.4,2.5-2.5);          
     \draw [ultra thick] (2.4,2.5-2.5) -- (2.8,2.5-2.5);    
     \draw [ultra thick] (1.4,3.5-2.5) -- (1.8,3.5-2.5);     
     \draw [ultra thick] (2,3.5-2.5) -- (2.4,3.5-2.5);     
     \draw [ultra thick,dashed,blue,->,>=stealth] (1.1,2.6-2.5) -- (1.5,3.4-2.5);  
     \draw [ultra thick,dashed,blue,<-,>=stealth] (1.3,2.6-2.5) -- (1.7,3.4-2.5);

     \draw [ultra thick] (4,2.5-2.5) -- (4.4,2.5-2.5);          
     \draw [ultra thick] (5.4,2.5-2.5) -- (5.8,2.5-2.5);    
     \draw [ultra thick] (4.4,3.5-2.5) -- (4.8,3.5-2.5);     
     \draw [ultra thick] (5,3.5-2.5) -- (5.4,3.5-2.5);     
     
     \draw [dashed,blue,->,>=stealth] (4.1,2.6-2.5) -- (5.1,3.4-2.5); 
     \draw [dashed,blue,<-,>=stealth] (4.3,2.6-2.5) -- (5.3,3.4-2.5); 

     \node at (0,2-2.5) {$T^{\alpha\beta,\pm}_{aa}:$};     
     \node at (2,2-2.5) {$V^\alpha_{ac}V^\beta_{ca}$};     
     \node at (3.45,2-2.5) {$+$};     
     \node at (5,2-2.5) {$V^\alpha_{ad}V^\beta_{da}$};
     
     \node at (0.8,2.5-2.5) {$a$};
     \node at (3,2.5-2.5) {$b$};
     \node at (1,3.5-2.5) {$c$};
     \node at (2.6,3.5-2.5) {$d$};
     
     \node at (0.8+3,2.5-2.5) {$a$};
     \node at (3+3,2.5-2.5) {$b$};
     \node at (1+3,3.5-2.5) {$c$};
     \node at (2.6+3,3.5-2.5) {$d$};

     \node at (0,3.5) {A)};
     \node at (0,1) {B)};
    
    \end{tikzpicture}
\end{center}
 \caption{Diagrammatic representation of the virtual transitions contributing to Raman relaxation, in panel A), and fourth-order pure dephasing, in panel B). Arrows intensity is proportional to rate of the transitions.}
 \label{Fig4}
\end{figure}

Fig. \ref{Fig3} also shows that, at the second order, $T_2$ and $T_1$, are identical. This is consistent with a vanishing $(T_2^{*})^{-1}$, being a zero-energy process ($\omega=0$) not achievable by exchanging a single phonon at the time. At the fourth order, we instead observe that $T^*_2$ is the limiting factor for $T_2$. The large difference between $\tau$ and $T_2$ result can understood by studying the terms of Eq. \ref{R4T}. In the case of $\tau$, the relevant jump operators' matrix elements are $L^{\kappa}_{ba}$, which at the fourth order include contributions from $T^{\alpha\beta,\pm}_{ba}$. The latter is dominated by virtual transitions to the firist excited KD\cite{lunghi2022toward} mediated by the product of matrix elements $V^\alpha_{bc}V^\beta_{ca}$, as schematically represented in Fig. \ref{Fig4}A. The electronic structure of SMMs such DyCp is engineered specifically to make $V_{ca}$ very small when $c$ and $a$ are sampled from opposite sides of the magnetization reversal barrier, making magnetic relaxation progressively more inefficient as the KDs get closer to pure $M_J$ states. In the case of fourth-order pure dephasing, however, the relevant terms of jump operators are of the form $L^{\kappa}_{aa}$ and $T^{\alpha\beta,\pm}_{aa}$, which instead include the product of matrix elements $V^\alpha_{ac}V^\beta_{ca}$ (see Fig. \ref{Fig4}B). While this number will be vanishingly small when $a$ and $c$ correspond to states with $M_J$ of opposite sign, it will be large for transitions to a KD with $\Delta M_J=\pm 1$, explaining the efficiency of Raman pure dephasing. Finally, we note that the fourth-order contribution to $T_2$ follows the same $T^{-3}$ power law trend as $\tau$, as is also mediated by THz optical phonons\cite{lunghi2022toward}.

\textit{Discussion.} Here we have extended the treatment of time-local master equations to the simulation of coherence terms up to the fourth order. These explicit expressions are novel, to the best of our knowledge, and apply generally to any open quantum system coupled to a bosonic harmonic bath. The role of phonons to pure dephasing has been overlooked under the general assumption that only dipolar interactions have a contribution. Only recently, second-order spin-phonon contributions to $T_2$ have been theoretically predicted in spin-1/2 molecules\cite{garlatti2023critical} and solid-state defects\cite{mondal2023spin}, and later experimentally observed\cite{han2025solid}, illustrating for the first time that spin-phonon coupling contributes to pure dephasing. Here, we have individuated a new phonon pure-dephasing mechanism that arises at the 4th order of theory and is extremely efficient in magnetic anisotropic systems, undercutting the befit of having a large magnetic anisotropy by about orders of magnitude. Reproducing the present results in experiments is challenging due to the vanishing matrix elements of a transverse magnetic field among these states, but continuous-wave electron paramagnetic resonance spectra for such highly axial and anisotropic Dy SMMs have nonetheless been reported\cite{parmar2025direct}, suggesting that our numerical predictions can be experimentally verified. In addition, the present results are completely general, and we expect any spin system affected by two-phonon processes to present such phenomenology, albeit not to the same extent of DyCp, thus greatly expanding the pool of compounds that can be used to explore phononic pure dephasing. Based on Eq. \ref{R4T}, future progress in increasing spin coherence in molecular systems should come from increasing the energy of the lowest-lying KDs and reducing spin-phonon coupling. In this respect, spin-1/2 coordination compounds and organic radicals are most promising, naturally presenting high-energy electronic excitations\cite{mariano2025role,gorgon2023reversible}, and work in this direction is ongoing. Alternative strategies are the exploration of high-frequency dynamical decoupling\cite{han2025solid} or the use of the multi-level structure of SMMs to engineer quantum states resilient to phonon decoherence\cite{ratini2025mitigating}.

\vspace{0.2cm}
\noindent
\textbf{Acknowledgements and Funding}\\
This project has received funding from the European Research Council (ERC) under the European Union’s Horizon 2020 research and innovation programme (grant agreement No. [948493]). Computational resources were provided by the Trinity College Research IT and the Irish Centre for High-End Computing (ICHEC).

\vspace{0.2cm}
\noindent
\textbf{Conflict of interests}\\
The authors declare no competing interests.


%

\end{document}